\documentclass[prl,aps,amssymb,twocolumn,showpacs,floatfix]{revtex4}

\usepackage{graphics}
\usepackage{epsfig} 

\usepackage{dcolumn}

\usepackage{amsmath}

\begin{document}

\newcommand\dd{{\operatorname{d}}}
\newcommand\sgn{{\operatorname{sgn}}}
\def\Eq#1{{Eq.~(\ref{#1})}}
\def\Ref#1{(\ref{#1})}
\newcommand\e{{\mathrm e}}
\newcommand\cum[1]{  {\Bigl< \!\! \Bigl< {#1} \Bigr>\!\!\Bigr>}}
\newcommand\vf{v_{_\text{F}}}
\newcommand\pf{p_{_\text{F}}}
\newcommand\ef{{\varepsilon} _{\text{\sc f}}}
\newcommand\zf{z_{_\text{F}}}
\newcommand\zfi[1]{{z_{_\text{F}}}_{#1}}
\newcommand\av[1]{\left<{#1}\right>}
\def\det{{\mathrm{det}}}
\def\Tr{{\mathrm{Tr}}}
\def\Li{{\mathrm{Li}}}
\def\tr{{\mathrm{tr}}}
\def\sgn{{\mathrm{sgn}}}
\def\im{{\mathrm{Im}}}

\title{Bosonization out of equilibrium}

\author{D. B. Gutman$^{1,2}$, Yuval Gefen$^3$, and A. D. Mirlin$^{4,1,2,5}$}
\affiliation{\mbox{$^1$Institut f\"ur Theorie der kondensierten Materie,
 Universit\"at Karlsruhe, 76128 Karlsruhe, Germany}\\
\mbox{$^2$DFG Center for Functional Nanostructures,
 Universit\"at Karlsruhe, 76128 Karlsruhe, Germany}\\
\mbox{$^3$Dept. of Condensed Matter Physics, Weizmann Institute of
  Science, Rehovot 76100, Israel}\\
\mbox{$^4$Institut f\"ur Nanotechnologie, Forschungszentrum Karlsruhe,
 76021 Karlsruhe, Germany}\\
\mbox{$^5$Petersburg Nuclear Physics Institute, 188300 St.~Petersburg, Russia}
}

\date{\today}

\begin{abstract}

We develop a bosonization technique for one-dimensional fermions out
of equilibrium. The approach is used to study a quantum wire
attached to two electrodes with arbitrary energy distributions. The
non-equilibrium electron Green function is expressed in terms of
functional determinants of a single-particle``counting'' operator
with a time-dependent scattering phase. The result reveals an
intrinsic relation of dephasing and energy redistribution in
Luttinger-liquids to ``fractionalization'' of electron-hole
excitations in the tunneling process and at boundaries with leads.

\end{abstract}



\maketitle

One-dimensional (1D) interacting fermionic systems are described as
Luttinger liquids (LL) \cite{gnt}, whose  experimental
manifestations include   carbon nanotubes,  semiconductor, metallic
and polymer nanowires, as well as quantum Hall edges. While
equilibrium LL have been extensively explored, there is currently a
growing interest in {\it non-equilibrium} phenomena on the
nanoscale. The major obstacle for a theoretical study of quantum
wires out of equilibrium is that bosonization, a technique most
suitable for analytic treatment of LL, has been so far restricted to
equilibrium situation.

\begin{figure}
\includegraphics[width=\columnwidth,angle=0]{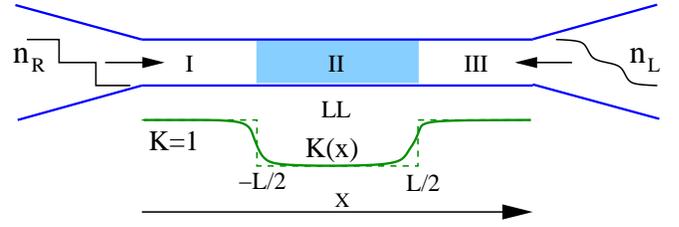}
\caption{Schematic view of a LL conductor connected to leads
with two different incoming fermionic distributions. The LL
interaction parameter $K(x)$ is also shown; the dashed line
corresponds to its sharp variation at the boundaries.}
\label{fig1}
\vspace*{-0.3cm}
\end{figure}

In this work we  develop  a bosonic theory for interacting 1D fermions
under generic non-equilibrium conditions.
The configuration we have in mind is shown in Fig.\ref{fig1}:
electrons with distributions $n_\eta(\epsilon)$ ($\eta=R,L$ labels
right- and left-movers) are injected into the LL wire from two
non-interacting electrodes. A particularly interesting case involves
double-step distributions \cite{GGM_2008}.  Our goal is to
calculate  the electron Green functions (GFs) 
$G^\gtrless(\tau)$ measurable in  tunneling spectroscopy experiments 
\cite{Birge,GGM_2009}. 

Our microscopic description of the problem
begins with Keldysh action in terms of fermionic fields $\psi(t,x)$ 
\cite{Kamenev},
\begin{eqnarray}&&
\label{TL}
S=S_0[\psi]+S_{\rm ee}[\psi]\,,\nonumber \\&&
S_0[\psi]=i\sum_{\eta}\int_c dt \int dx \psi^\dagger_\eta
\partial_\eta\psi_\eta \,, \nonumber \\&&
S_{\rm ee}[\psi]=-\sum_\eta\int_c dt \int dx g(x) (\rho_{\eta}\rho_{-\eta}+
\rho_{\eta}\rho_{\eta})\,.
\end{eqnarray}
Here $\rho_\eta=\psi^\dagger_\eta\psi_\eta$ are density fields,
$\partial_{\eta} = \partial_t+\eta v\partial_x$, $v$ is the Fermi
velocity,  $g(x)$ is a spatially dependent interaction strength, and
 $\eta=\pm1$ stands for right/left moving electrons.
%
The essence of bosonization is a reformulation of the theory in
terms of bosonic (density) fields. The interacting part of the
action $S_{ee}$ is already expressed in terms of density modes
$\rho_\eta$. The challenge is to bosonize the free part of the
action, where  information concerning the state of the
non-interacting fermionic system is encoded. The bosonic counterpart
of the free part of the action is found by requiring that it
reproduces correlation functions of non-equilibrium free Fermi gas
\cite{unpublished},
\begin{eqnarray}&&
\label{action}
S_{0} =\sum_\eta
(\rho_{\eta}
\Pi_\eta^{a^{-1}}
\bar{\rho}_{\eta}-i\ln Z_\eta[\bar{\chi}_\eta]).
\end{eqnarray}
Here we have performed a rotation in Keldysh space, decomposing
fields into classical and quantum components, $\rho,\bar{\rho}=
(\rho_+\pm\rho_-)/\sqrt{2}$, where the indices $+$ and $-$ refer to
the two branches of the Keldysh contour; $\Pi_\eta^a$ is the
advanced component of polarization operator, and
$Z_\eta[\bar{\chi_\eta}]$ is a partition function of free 
fermions moving in the field
\begin{equation}
\label{c1}
\bar{\chi}_\eta=\Pi_\eta^{a^{-1}}\bar{\rho}_{\eta}.
\end{equation}

Expansion of $Z_\eta[\bar{\chi_\eta}]$ in $\bar{\chi_\eta}(t,x)$
generates an infinite series, $i\ln
Z_\eta[\bar{\chi_\eta}]=\sum_n(-1)^{n+1}\bar{\chi_\eta}^n{\cal
  S}_{\eta n}/n$,
governed by irreducible fermionic  density correlation functions,
${\cal S}_{\eta n}(t_1,x_1;\ldots;t_n,x_n) \equiv
\langle\langle\rho_{\eta 1}\rho_2 \dots \rho_{\eta
n}\rangle\rangle$, representing cumulants of quantum noise
\cite{Levitov-noise}. At equilibrium the random phase approximation
(RPA) is exact \cite{DzLar:73}, ${\cal S}_{\eta n}=0$ for $n> 2$,
and one obtains the Gaussian theory of conventional bosonization. In
a generic non-equilibrium situation, the bosonic action contains
terms of all orders with no small parameter; the idea to proceed
analytically in a controlled  manner may seem hopeless. This,
however,  is not the case: non-equilibrium bosonization is an
efficient framework in which the functional integration can be
performed exactly. Two observations are crucial here. First,
$Z_\eta$ depends only on the quantum component $\bar{\rho}$, so that
the   action, Eq.(\ref{action}), is linear with respect  to the
classical component $\rho$ of the density field. Hence the
integration with respect to $\rho$ can be performed exactly,
generating an equation that fixes $\bar{\rho}$. As a result, the
many-body field theory is mapped onto a problem which has a level of
complexity of quantum mechanics. Second, all vertex functions ${\cal
S}_{\eta n}$ determining $Z_\eta[\bar{\rho_\eta}]$ are restricted to
the mass shell ($\omega=\eta vq$) with respect to all of their
arguments. This will render the effective ``quantum mechanics''
one-dimensional.

We apply this formalism first to free fermions away from
equilibrium, and obtain the single-particle GFs
$G^<_{0,\eta}(x,t)=i\langle
\psi_\eta^\dagger(0,0)\psi_\eta(x,t)\rangle$,
$G^>_{0,\eta}(x,t)=-i\langle
\psi_\eta(x,t)\psi^\dagger_\eta(0,0)\rangle$  by means of the
bosonic theory. Since $G_{0,\eta}$ depends on $\tau=t-\eta x/v$
only, we set $x=0$ and $t=\tau$ in the arguments of $G_{0,\eta}$.
The fermionic operators are expressed in terms of the bosonic field
in the standard way, $\hat{\psi}_{\eta}(x)\simeq(\Lambda/2\pi
v)^{1/2}e^{\eta ik_Fx} e^{i\hat{\phi}_\eta(x)}$, where
$\hat{\rho}_{\eta}(x)=\eta\partial_x \hat{\phi}_{\eta}/2\pi$ and
$\Lambda$ is an  ultraviolet cut-off. After integration over the
classical component we find that quantum component $\bar{\rho}$
satisfies a continuity equation
\begin{equation}
\label{d1}
\partial_t\bar{\rho}_\eta+\eta v\partial_{x}\bar{\rho}_\eta= j(t,x)\,
\end{equation}
with  the source term
$j= \delta(x)[\delta(t-\tau)-\delta(t)]/\sqrt{2}$. The fermionic GF 
is obtained as [with $\bar{\chi}_\eta$ given by Eq.(\ref{c1})] 
\begin{equation}
\label{a1}
G_{0,\eta}^\gtrless(\tau)=-\frac{1}{2\pi v}\frac{1}{\tau\mp i/\Lambda}
Z_\eta[\bar{\chi}_\eta]\:.
\end{equation}
The mass-shell
nature of $S_{\eta n}$ implies that $Z_\eta[\bar{\chi}_\eta]$
depends only on the world-line integral
\begin{equation}
\label{d1a}
\delta_\eta(t)=\sqrt{2} \int_{-\infty}^\infty
d\tilde{t}\: \bar{\chi}_\eta(\tilde{t}-t,\eta v\tilde{t}).
\end{equation}
  More specifically, we find \cite{unpublished}
\begin{equation}
\label{a2}
Z_\eta[\bar{\chi}_\eta] =
\det[1
  + (e^{-i\hat{\delta_\eta}}-1)\hat{n}_\eta]
\equiv \Delta_\eta[\delta_\eta(t)].
\end{equation}
For the free-fermion problem the phase
$\delta_\eta(t)=\lambda\omega_\tau(t,0)$
where $w_\tau(t,\tilde{t})=\theta(\tilde{t}-t)-\theta(\tilde{t}-t-\tau)$
is a ``window function'' and
$\lambda=2\pi$. Thus, $Z_\eta[\bar{\chi}_\eta] =
\Delta_{\eta\tau}(2\pi)$, where $\Delta_{\eta\tau}(\lambda)$ is the
determinant (\ref{a2}) for a rectangular pulse \cite{footnote-normalization}.

Since $\hat{n}_\eta$ is diagonal in energy space, while
$\hat{\delta_\eta}$ is diagonal in time space, they do not commute,
making the determinant non-trivial. Determinants of the type
(\ref{a2}) have appeared  in a theory of counting statistics
\cite{Levitov-noise}. 
Specifically, the generating function of current
fluctuations 
$\kappa(\lambda)=\sum_{n=-\infty}^\infty e^{i n\lambda}p_n$ (where $p_n$
is the  probability of  $n$ electrons being transferred through the
system in a given time window $\tau$) has the same structure as
$\Delta_{\eta\tau}(\lambda)$. Taylor expansion of
$\ln\kappa(\lambda)$ around $\lambda=0$ defines cumulants of current
fluctuations.

\begin{figure}
\includegraphics[width=0.8\columnwidth,angle=0]{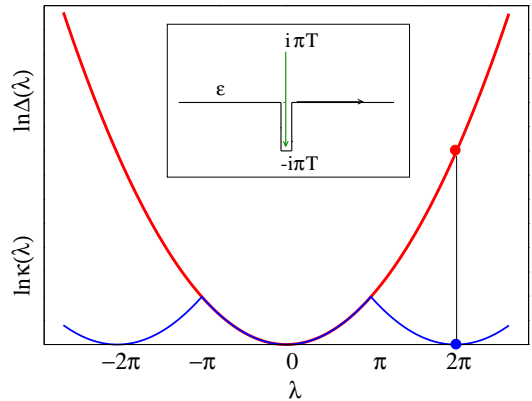}
\caption{Analytic $\Delta_{\eta\tau} (\lambda)$ vs. periodic 
  $\kappa(\lambda)$ continuation of the functional determinant. The value of
  $\Delta_{\eta\tau}$ at  $\lambda=2\pi$ determines the free-electron
  GF, while  $\ln\kappa(2\pi)=0$ in view of charge quantization.
As an example, the equilibrium case is shown.
 {\it 
    Inset:} contour of integration for the quasiclassical limit,
  Eq.~(\ref{a5}),  of
  $\Delta_{\eta\tau}(\lambda)$ is deformed, since a singularity of the integrand
  crosses the real axis at $\lambda=\pi$.}
\label{fig5a}
\vspace*{-0.3cm}
\end{figure}

According to its definition, $\kappa(\lambda)$ is $2\pi$-periodic,
which is a manifestation of  charge quantization. Thus,
$\kappa(2\pi)\equiv 1$ is trivial. On the other hand, we have found
that the free electron GF is determined by the value of
the functional determinant exactly at  $\lambda=2\pi$. A resolution
of this apparent paradox is as follows: the determinant
$\Delta_{\eta\tau}(\lambda)$ should be understood as an analytic
function of $\lambda$ increasing from 0 to $2\pi$. On the other
hand, $\kappa(\lambda)$ is non-analytic at the branching points
$\lambda=\pm\pi, \pm 3\pi,\ldots$. To demonstrate this, it is
instructive to consider the equilibrium case (temperature $T$). 
Then the expansion of
$\ln\Delta_{\eta\tau}(\lambda)$ in $\lambda$ is restricted to the
$\lambda^2$ term (since RPA is exact). It is easy to check that the
$\lambda=2\pi$ point on this parabolic dependence correctly
reproduces the fermion GF via Eqs.~(\ref{a1}),
(\ref{a2}). As to the counting statistics $\ln\kappa(\lambda)$, it
is quadratic only in the interval $[-\pi,\pi]$ and is periodically
continued beyond this interval, see Fig.~\ref{fig5a}.

The difference in analytical properties of $\kappa(\lambda)$ and
$\Delta_{\eta\tau}(\lambda)$ becomes especially transparent if one studies
the semiclassical (long-$\tau$) limit,
\begin{equation}
\label{a5}
\ln \Delta_{\eta\tau}[\lambda]=\frac{\tau}{2\pi\hbar}\int d\epsilon
\ln[1+(e^{-i\lambda}-1) n_\eta(\epsilon)]\,.
\end{equation}
For small $\lambda$ the singularity of the integrand closest to the
real axis is located near $\epsilon = i\pi T$. As $\lambda$
increases, the singularity moves towards the real axis, crosses it
at $\lambda=\pi$ and finally approaches $\epsilon=-i\pi T$ as
$\lambda\to2\pi$ (see  inset of Fig.\ref{fig5a}). The integral for
$\ln\kappa(\lambda)$ is taken along the real axis, resulting in
non-analyticity at $\lambda=\pi$ and in zero value at
$\lambda=2\pi$. On the other hand, the contour for
$\ln\Delta_{\eta\tau}(\lambda)$ is deformed to preserve analyticity,
resulting in the long-$\tau$ asymptotics $\ln\Delta_{\eta\tau} =
-\tau T\lambda^2/4\pi$.

The value of $\lambda=2\pi$ appearing in the bosonic representation
of the free-fermion GF $G_{0,\eta}(\tau)$ has a clear
physical meaning: a fermion is a $2\pi$-soliton in the bosonic
formalism. A natural question to ask is whether  values of
$\Delta_{\eta\tau}(\lambda)$ away from $\lambda=2\pi$ are physically
important. To see that this is indeed the case, consider  the Fermi
edge singularity (FES) problem. In this problem, an electron excited
into the conduction band, leaves behind a localized hole, resulting
in an $s$-wave scattering phase shift, $\delta_0$, of the
conducting electrons \cite{Nozieres}. In the mesoscopic context, the
FES manifests itself in resonant tunneling experiments \cite{Geim}.
The problem is effectively described by chiral 1D electrons
interacting with a core hole that is instantly ``switched on''. As
was shown in \cite{Schotte}, taking into account the core hole in
the bosonization approach amounts to a replacement of
$e^{i\hat{\phi}}$ by $e^{i(1-\delta_0/\pi)\hat{\phi}}$ in the boson
representation of the fermionic operator. Within our non-equilibrium
formalism, this implies a replacement $j\rightarrow
(1-\delta_0/\pi)j$ in Eq.~(\ref{d1}). Performing the derivation as
in the free-fermion case, we thus obtain the non-equilibrium FES
GF for electrons with an arbitrary distribution
$n(\epsilon)$,
\begin{equation}
\label{b1}
G^\gtrless(\tau)=\mp i\Lambda\Delta_{\tau}[2(\pi-\delta_0)] /
2\pi v (1\pm i\Lambda\tau)^{(1-\delta_0/\pi)^2}.
\end{equation}
It is easy to check that at equilibrium this reproduces known
results \cite{Nozieres,Schotte}. For a non-equilibrium double-step
distribution the FES problem was studied in
\cite{Abanin} within a complementary (fermionic) approach 
(separating of $G(\tau)$ into a product of open line and closed loop
contributions) bearing an analogy with functional bosonization
\cite{Fogedby:76}. While both approaches yield equivalent
results, this equivalence is highly non-trivial \cite{unpublished}.
Functional determinants of the type $\Delta_\tau(\lambda)$ can be
efficiently evaluated numerically \cite{Nazarov}. 

We are now prepared to address  the problem formulated in the
beginning of the paper: an interacting quantum wire out of
equilibrium, Fig.\ref{fig1}. We consider $G_R^\gtrless(\tau)$ 
for the
tunneling point ($x=0$) located inside the interacting part;
generalization to tunneling into a non-interacting region is
straightforward. Integrating over the classical components
$\rho_\eta$ as explained above, we obtain equations satisfied by the
quantum components $\bar{\rho}_\eta$ of the density fields,
\begin{eqnarray}&&
\label{a7}
\partial_t\bar{\rho}_{R}+\partial_{x}\left
[(v+{g\over 2\pi})\bar{\rho}_{R}+\frac{g}{2\pi}\bar{\rho}_{L}\right]=j
\nonumber \\&&
\partial_t\bar{\rho}_{L}-\partial_{x}\left[(v+\frac{g}{2\pi})\bar{\rho}_{L}+
\frac{g}{2\pi}
\bar{\rho}_{R}\right]=0\,.
\end{eqnarray}
The solution of Eq.~(\ref{a7}) determines the phases
$\delta_\eta(t)$ according to Eqs.~(\ref{d1a}), (\ref{c1}), which
can be cast in the form
\begin{equation}
\label{a7a}
\delta_\eta(t)=
-2\pi\sqrt{2}\:
\eta\lim_{\tilde{t}\rightarrow-\infty}\int_0^{\eta v(\tilde{t}+t)}d\tilde{x}\:
\bar{\rho}_\eta(\tilde{x},\tilde{t})\,.
\end{equation}
Remarkably, Eq.~(\ref{a7a}) expresses the phase $\delta_\eta(t)$
affected by the electron-electron interaction,  through the
asymptotic behavior of $\bar{\rho}(x,t)$ in the non-interacting
parts of the wire (regions I and III in Fig.\ref{fig1}). The phases
$\delta_\eta(t)$ determine the GFs via
\begin{equation}
\label{d3}
G^\gtrless_R(\tau)=-\frac{\Delta_R[\delta_R(t)]\Delta_L[\delta_L(t)]}
{2\pi v (\pm i\Lambda)^\gamma(\tau\mp i/\Lambda)^{1+\gamma}},
\end{equation}
where $\gamma=(1-K)^2/2K$ and $K=(1+g/\pi v)^{-1/2}$ is the standard LL
parameter in the interacting region.

\begin{figure}
\includegraphics[width=0.95\columnwidth,angle=0]{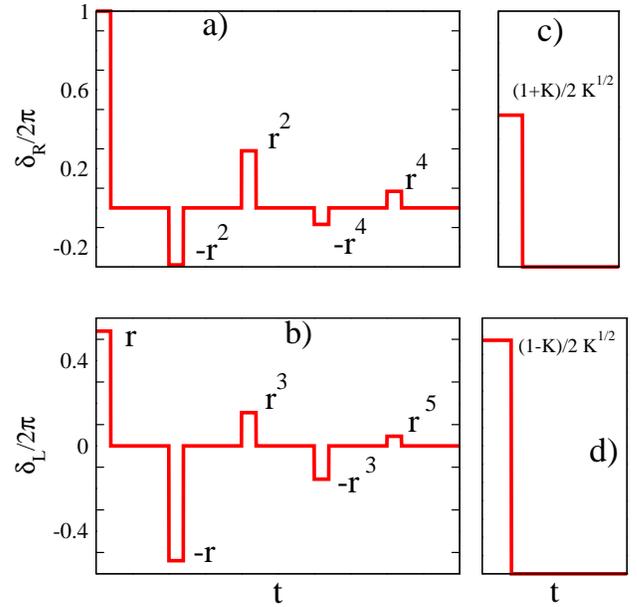}
\caption{Phases $\delta_\eta(t)$ entering Eq.~(\ref{d3})
 for the GFs for sharp [a),b); $r=(1-K)/(1+K)$] and adiabatic [c),d)]
  boundaries. }
\label{fig2}
\vspace*{-0.3cm}
\end{figure}

To explicitly evaluate $\delta_\eta(t)$ for the structure of
Fig.~\ref{fig1}, it is convenient to rewrite Eqs. (\ref{a7}) as a
second-order differential equation for the current
$\bar{J}=v(\bar{\rho}_R-\bar{\rho}_L)$,
\begin{equation}
\label{d2}
(\omega^2+\partial_x
u^2(x)\partial_x)J(\omega,x)=0\,,\ \  x \neq 0,
\end{equation}
where $u(x)=v(1+g(x)/\pi v)^{1/2}$ is a spatially dependent plasmon
velocity. Reflection and transmission of plasmons on both boundaries
is characterized by the coefficients $r_\eta$, $t_\eta$
($r_\eta^2+t_\eta^2 =1$); for simplicity, we assume them to be
constant over a characteristic frequency range  $\omega \sim
\tau^{-1}$. Subsequently  $\delta_\eta(t)$ is found to be a
 superposition of rectangular pulses,
$\delta_\eta(t)=\sum_{n=0}^\infty \delta_{\eta,n}w_\tau(t,t_n)$,
where $t_n=(n+1/2-1/2K)L/u$, and
\begin{eqnarray}&&
\delta_{\eta,2m}=\pi t_\eta r_L^mr_R^m(1+\eta K)/\sqrt{K}\,,
\nonumber \\&&
\delta_{\eta,2m+1}=
-\pi t_\eta r_\eta^m r_{-\eta}^{m+1} (1-\eta K)/\sqrt{K}\,.
\end{eqnarray}
%
%
The phases $\delta_\eta(t)$ are shown in Fig.~\ref{fig2} for two
limits of adiabatic ($r_\eta=0$) and sharp [$r_\eta=(1-K)/(1+K)$]
boundaries.

In physical terms,  $\delta_\eta(t)$ characterizes phase
fluctuations in the leads that arrive at the time interval
$[0,\tau]$ at the measurement point $x=0$,  governing the
dephasing and the energy distribution of electrons encoded in the
GFs $G^\gtrless_\eta(\tau)$. Up to inversion of time, 
one can think of $\delta_\eta(t)$
as describing the fractionalization of a phase pulse (electron-hole
pair) injected into the wire at point $x$ during the time interval
$[0,\tau]$. This is closely related to the physics of charge
fractionalization discussed earlier \cite{Maslov,lehur}. At the
first step, the pulse splits into two with relative amplitudes
$(1+K)/2$ and $(1-K)/2$ carried by plasmons in opposite directions,
cf. Ref.~\cite{lehur}. 
As each of these pulses reaches the corresponding boundary, another
fractionalization process takes place: a part of it is transmitted
into a lead, while the rest is reflected. The reflected pulse
reaches the other boundary, is again fractionalized there, etc. 
Let us stress an
important difference between boundary fractionalization of transmitted charge
\cite{Maslov} and that of dipole pulses. While
in the former case the boundaries can
always be thought as sharp, in the present problem the way $K(x)$ 
is turned on is crucially important for
reflection coefficients $r_\eta$ at $\omega\sim\tau^{-1}$.

For $\tau \ll L/u$
the coherence of plasmon scattering  may  be  neglected and
the result splits into a product
\begin{equation}
\label{c10}
\Delta_\eta[\delta_\eta(t)]\simeq\prod_{n=0}^\infty
\Delta_{\eta\tau}(\delta_{\eta,n})
\,,
\end{equation}
with each factor  representing a contribution of
a single phase pulse
$\delta_{\eta,n}(t)=\delta_{\eta,n}w_\tau(t,0)$.
For the ``partial equilibrium'' state (when $n_R(t)$ and $n_L(t)$
are of Fermi-Dirac form but with different temperatures) the
functional determinants are gaussian functions of phases, 
reproducing  earlier results of functional bosonization
\cite{GGM_2009}.

\begin{figure}
\includegraphics[width=0.95\columnwidth,angle=0]{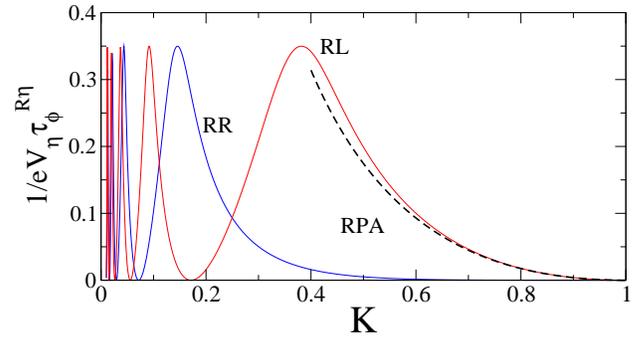}
\caption{Dephasing rates $\tau_\phi^{RR}$ and $\tau_\phi^{RL}$ as
function of LL parameter $K$ for the adiabatic case and double step
distributions with $a_R=a_L=1/3$. }
\label{fig5}
\vspace*{-0.7cm}
\end{figure}

We now apply our general result (\ref{c10}) to the ``full
non-equilibrium'' case, when $n_\eta(\epsilon)$ have a double step
form,  $n_\eta(\epsilon)=
a_\eta\theta(V_\eta-\epsilon)+(1-a_\eta)\theta(-\epsilon)$. 
 The split zero-bias-anomaly dips are then broadened by
non-equilibrium dephasing rate, $1/\tau_\phi^\eta$
\cite{GGM_2008,GGM_2009}. 
The latter can
be found from the long-time asymptotics of $G_\eta(\tau)$, employing
Eq.~(\ref{a5}). We obtain
$1/\tau_\phi^R=1/\tau_\phi^{RR}+1/\tau_\phi^{RL}$, where
$1/\tau_\phi^{\eta\eta'}$ is the contribution to dephasing of $\eta$
fermions governed by the distribution of $\eta'$ fermions. Focussing
on the adiabatic limit, we find these dephasing rates to be
\begin{equation}
\label{c11}
1/\tau_\phi^{R\eta}=-\frac{eV_\eta}{2\pi}\ln\left(
1-4a_\eta(1-a_\eta)\sin^2
\frac{\pi(1+\eta K}{2\sqrt{K}}\right),
\end{equation}
see Fig.~\ref{fig5}.
Two remarkable features of this result should be pointed out. First,
for certain values of the interaction parameter $K$ (different for
$\eta=R,L$) the dephasing rates $1/\tau_\phi^{R\eta}$ vanish.
Second, consider the weak-interaction regime,
$\gamma\ll 1$. We then obtain
$1/\tau_\phi^{RL}\simeq\pi\gamma eV_L a_L(1-a_L)$ and
$1/\tau_\phi^{RR}\simeq\pi(\gamma^2/8) eV_R a_R(1-a_R)$. This should
be contrasted with RPA which predicts equal $1/\tau_\phi^{RL}$ and
$1/\tau_\phi^{RR}$ \cite{GGM_2008}. While $1/\tau_\phi^{RL}$ agrees
with the RPA result,  $1/\tau_\phi^{RR}$ is parametrically smaller
(suppressed by an extra factor of $\gamma$). The reason for this
failure of  RPA is clear from our analysis. For a weak interaction
the contributions of R and L movers to $G_R$ are given by the
functional determinants $\Delta_{\eta\tau}(\delta_\eta)$ with phases
(for an adiabatic barrier) $\delta_L\simeq (1-K)\pi$ and $\delta_R
\simeq \pi(1+K)$. While the contribution of the small phase
$\delta_L$ is captured correctly by RPA, a small-$\delta$ expansion
of $\ln\Delta_{R\tau}(\delta_R)$ fails for large $\delta_R$ (apart
from equilibrium and ``partial equilibrium'' where
$\ln\Delta_{\eta\tau}(\delta)\propto \delta^2$.)

To conclude, we have developed a non-equilibrium bosonization technique
and used it to find GFs for tunneling
spectroscopy of a LL conductor.  Our solution  is presented in
terms of functional determinants of single particle operators. It
elucidates the relation between the charge
fractionalization on one hand, and energy redistribution and
non-equilibrium dephasing on the other hand. We have calculated 
dephasing broadening of the ZBA dips for double-step
distributions. Our approach has various further applications,
including spinful electrons, interferometry, noise, and many-body
entanglement \cite{unpublished}.  

We thank D.~Abanin, D.~Bagrets, L.~Glazman,
L.~Levitov, P.~Ostrovsky, and Y.~Nazarov for discussions. 
This work was supported
by GIF, Einstein Minerva
Center, US-Israel BSF, ISF,
Minerva Foundation, and DFG SPP 1285.


\begin{thebibliography}{99}
\bibitem{footnote-normalization} We normalize the determinants
  $\Delta_{\eta\tau}(\lambda)$ to their equilibrium, zero-temperature value.
\end{thebibliography}
\end{document}